# MLOps: A Step Forward to Enterprise Machine Learning

Abdullah Ikram Ullah Tabassam
*Department of Computer Science and Mathematics*
*Liverpool John Moores University*
Liverpool, United Kingdom
abdullahdar2017@gmail.com
csmaulla@ljmu.ac.uk

*Abstract*— Machine Learning Operations (MLOps) is becoming a highly crucial part of businesses looking to capitalize on the benefits of AI and ML models. This research presents a detailed review of MLOps, its benefits, difficulties, evolutions, and important underlying technologies such as MLOps frameworks, Docker, GitHub actions, and Kubernetes. The MLOps workflow, which includes model design, deployment, and operations, is explained in detail along with the various tools necessary for both model and data exploration and deployment. This article also puts light on the end-to-end production of ML projects using various maturity levels of automated pipelines, with the least at no automation at all and the highest with complete CI/CD and CT capabilities. Furthermore, a detailed example of an enterprise-level MLOps project for an object detection service is used to explain the workflow of the technology in a real-world scenario. For this purpose, a web application hosting a pre-trained model from TensorFlow 2 Model Zoo is packaged and deployed to the internet making sure that the system is scalable, reliable, and optimized for deployment at an enterprise level.

## I. Introduction

Over the last two decades, there has been a lot of progress in the field of Artificial Intelligence (AI), especially Machine Learning (ML). Researchers have been working on solving a huge range of complex problems using various methodologies and architectures based on Machine Learning, but these were all based on academics and research purposes only, for a long time. With the amount of advancement that is observable in this field, and availability to huge amounts of data, in the recent years, there is a lot of demand for ML solutions for business problems [1]. Many companies are switching their business to support AI. At this stage, the need of the hour for ML practitioners, is to take a step further from working in isolation, to creating an accessible and collaborative environment which is reproduceable at the same time so that it can be put to mass production and also monitored over time [1].

To create the said environment, the use of Machine Learning Operations (MLOps) [3] is necessary. To understand about MLOps in general and its importance in the production field, we divide the whole process into three critical stages. The first stage the Experimentation stage, where the data is collected, prepared for training the model, along with the model selection. Then comes the Development stage, in which the model is trained and tested. And the last is the Operations stage, where the model is hosted on servers as a service and is continuously retrained over new coming data and re-evaluated at the same time [4]. To define MLOps, it can be said that MLOps is now the standard way to deploy machine learning algorithms and manging their life cycle [5].

Deploying the ML algorithms to the servers for business use and industrializing the projects for advancement and betterment in various fields is one of the greatest advantages of MLOps but it also comes with its own challenges [6]. When the technologies are not much mature, the challenges tend to increase. At a low maturity level, only the traditional ML methods can be applied and a strong interaction between individual teams is very critical. The data engineers have to be in-touch with the data scientists' team, and they further have to be in-touch with the ML engineers and the Developers. The deviation from this results in a lot of technical issues [7]. A major challenge is to create robust pipelines [8] that are efficient as well. These pipelines need to work in harmony with each other, with same pre-processing libraries and dependencies so that it reduces the chances of errors to a minimum. Another challenge associated with MLOps is of versioning [9]. Version updates to ML services is not like code updates to basic software programs, as code updates can be made by just creating changes and testing them instantly, but for ML products, continuous retraining [10] and re-evaluation is a crucial step which consumes both time and resources. The continuous monitoring [11] of the service is also a big challenge for MLOps as there is always a chance to drift in the data and need to retrain the model, or worse, complete change the model to fit the latest trends and needs of the time. Keeping these challenges in mind, there is a need for selection of the appropriate tools necessary to provide better results as well as future monitoring and retraining of the model without disruption.

This paper is focused on discussing in detail, the importance, and evolution of MLOps and the tools and technologies related to the deployment and packaging of the ML pipelines as a business product. The coming sections will provide a complete workflow of the MLOps to provide a starting point for anyone looking to switch from just a basic ML model to a complete ML Business pipeline.

## II. Background

The term MLOps refers to the operations that are necessary in the lifecycle of an ML model [5]. MLOps is somewhat similar to DevOps [12], but has a huge difference at the same time. Before going deep into the similarities and differences, it is important to know what DevOps is and why is it necessary. DevOps emerged as an important field, somewhere around 2009 [13], and the main purpose was to provide workflows that could help increase the capability of software production by fast and automated end-to-end deployment [14][15][16][17]. The main purpose of DevOps is to create an environment where all the teams including Dev, test and Operations are linked together and work in harmony to produce better products increasing the speed and productivity at the same time because of automation. A typical DevOps workflow is shown in Figure *1*.

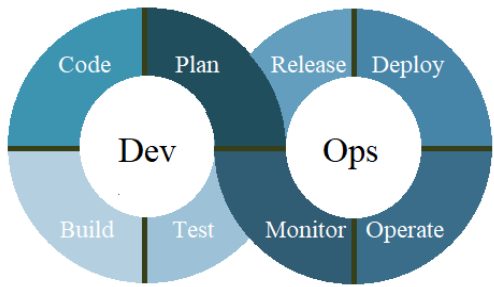

Figure 1: DevOps Workflow [18][19]

*A. DevOps vs. MLOps*

As discussed previously, DevOps creates a linkage between the developer tasks and operations task. To perform these task, two types of pipelines are used. One is called Continuous Integration (CI) pipeline and the other is called Continuous Delivery (CD) pipeline. The purpose of the CI pipeline is to add the feature of continuous integration of the program code by the developers to overcome bugs, and errors, and create small changes or improvements required to the code continuously after some interval of time. They keep on testing the software periodically, and based on the results of the test, they make these changes and integrate them to the previous code [20]. Similarly, after these changes to the code, they version the software and deliver the newer, better version to the end user through the CD pipeline so that the process of delivery is much faster and seamless [18]. Up till this point both DevOps and MLOps are same, but the difference starts from this point forward. The main difference lies in the fact that MLOps utilizes the principles of DevOps for ML model deployment instead of code. It uses both the CI and CD pipelines for continuous integration and delivery of the ML models [21] but the ML models versioning is not the same as code versioning. The Dev team improves the code, tests it and deploys it. But for ML models, ML code is improved, retrained with either the same, or in some cases new collected data. And then deployed to the server. This is why it needs another pipeline called the Continuous Training (CT) pipeline, in addition to the Continuous Integration – Continuous Delivery pipelines. This is the reason MLOps is usually considered more complex than DevOps [22][23][24]. The workflow of MLOps is represented in Figure *1* showing that in addition to Development and Operations, MLOps also deals with continuous retraining of the ML models on continuously increasing data.

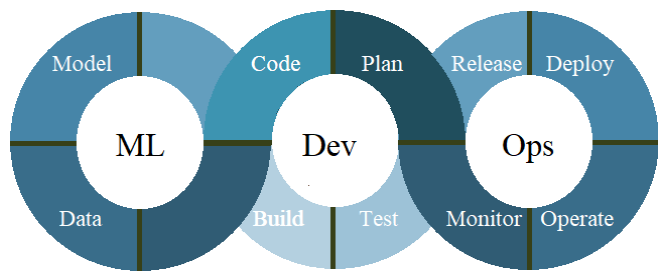

Figure 2: MLOps Workflow[24]

*B. ML Lifecycle*

This section discusses the phases through which the ML model goes through out its life cycle. The first phase is the data preparation phase. Numerous data sources are used for collecting the data and after necessary pre-processing and feature extraction, it is stored in the data stores. The data from the data stores is used for both model training and inferencing [32]. The feature store is beneficial as it allows to reuse the features without hassle of creating new features every time. It maintains the metadata so that there are no similar features with same definitions. Feature stores serve up to date features and help in using the same features for serving that were used in training. It must be made sure that the data has been stored properly following all the regulations by the authorities, keeping it anonymous and safe to use, throughout the process.

The next phase is the model development phase. In this phase, the best model for the task is selected and the data from the data store is used to train the model. Hyper-parameter tuning and evaluation is done keeping the industry standards and the application in mind. The choice of architecture for model development is important as the data is continuously changing and model must be maintained, so it is important to make sure that the best model that suit the task is used. Choosing a wrong model can cause a lot of challenges and can affect the performance of the model in many ways [26]. The data to be used, its features, the continuous iteration process [27] must be kept in mind. Version logging is also the part of this phase, and it helps to keep the model up to date [5]. The model also has to go through a lot of scrutiny both in the field of ethics and regulation. If it passes all the key performance criteria, it is pushed towards the next phase that is the deployment phase.

In the deployment phase, the model is deployed to the server using specialized tools. Making intelligent choice when selecting these tools is crucial. These choices are completely related to the application of the service and can be tasks like need of automation using CI/CD pipelines, using docker containers, version control or any other platform that could host the ML service [28].

After model deployment the most important part is the continuous monitoring of the model metrics as the ML models are highly susceptible to change and the data is always increasing. The data drift [5][29] can cause model performance to drop sharply. These factors can lead the model to become stale and the solution to this is to continuously monitor the model [30] and find out the main reason that caused the performance to drop. Retraining the model continuously on the new data helps to keep it up to date and increase the performance too.

*C. Evolution of MLOps pipelines – MLOps Maturity*

The evolution of MLOps systems is usually considered with the level of automation that they have. This is referred to as the maturity level of the model. There are various maturity levels depending on what level of automation they have [23]. The two main maturity models based on complexity and automation level are from Microsoft [31] and Google [32]. Microsoft has divided its model in five maturity levels. The first level has no automation at all. At the second level, it introduces DevOps but there is no MLOps. At the third level, the training is automated while the deployment is automated at the fourth level. Finally, level five introduces fully automated MLOps. Figure 3 shows the five levels of automation maturity in the Microsoft model. The model by Google divides the automation into three levels. The first level is called level 0 as it has no automation. Level 1, the second level, consists of automated ML pipeline, and at last maturity level, the CI-CD pipelines are automated. A deeper discussion on these levels will be done in the considerations section. Figure 4 shows the three layers of google maturity model.Figure 3

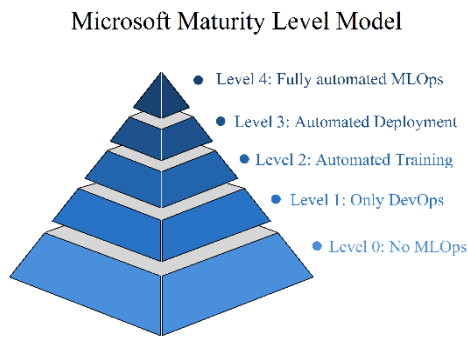

Figure 3: Microsoft maturity model [7][31]

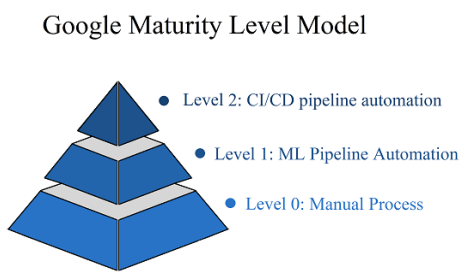

Figure 4: Google maturity model [7][31]

### D. Challenges and solutions

No doubt, using automated pipelines save time and resources, but these also have some challenges associated to them and the awareness of these challenges is necessary to cater them properly.

- **Compatible pipelines**

    MLOps uses various types of automated pipelines. The features are extracted in one pipeline and fed to another where the model is trained. The trained model is hosted using some other pipelines. Using these pipelines though increase the performance but it is important that all these pipelines are compatible to each other, and the choice of tools for automation is correct. If the pipelines are not compatible with each other or the pre-processing libraries do not match, it could cause many sorts of errors [7]. For example, Kerdo, SageMaker and MLFlow are brilliant tools for ML operations, but if not used with AWS or DataBricks, they only provide limited support for end IoT devices [33].

- **Monitoring and retraining**

    Monitoring is the most important part of MLOps as it helps increase the efficiency of the model, but it becomes a big challenge when there is a lot of data coming up all the time. This is the age of data and more and more data is adding up in the resources every day. It is important to ensure that the data we use to train/retrain our models is relevant and looks like how we expected it to be. Monitoring helps us see if the model neds to be re-trained on new data to increase the performance or if there is a data shift that would require us to completely change the model.

- **Overfitting and biased models**

    Overfitting is a real issue in the ML models. Evaluating models using the industry specified metrics and keep checking for overfitting can save a company from a lot of challenges associated to the performance. The model should perform in the real-world environment identical to the experimentation phase. In addition, chances of having biased models should be eliminated completely. A biased model can work completely fine and can be perfect with accordance to the technical terms, but the chances of a biased model to fail at ethical grounds are very high. These models can cause bad reputation for the business and can even lead to legal proceedings. The model should be tested thoroughly for negating ethical allegations before it is hosted on the server.

## III. CONSIDERATIONS

After discussing about what MLOps is, how it works and what are the challenges related to implementing MLOps on business models in an efficient way, we need to consider understanding some important things related to MLOps. These include the understanding of legal and ethical approach to MLOps as well as the tools and level of automation required for our model. This section is focused on providing some insights on the mentioned considerations in detail.

### A. Legal and ethical considerations

Before getting into MLOps, it is to be made sure that the models we serve at the internet are for the betterment of the people and cannot be biased towards a single group in anyway. On the contrary, it should treat everyone as equal. There has been a lot of uproar in the previous few years where AI models were found biased causing trouble to the masses. In addition to this, the privacy of the users should be taken seriously as the information they provide to the service are not supposed to be made public. It is to be made sure that the model does not save personal information like name, telephone numbers, addresses, bank account details etc in its data store, and even if it does for any task, it disposes this information off properly after the job is done. The laws like GDPR [34] and CCPA [36] are in place to make sure the model policies comply with the privacy policies set by the governments. Another thing to be considered is the use of proper licencing for all the tools and modules used to provide protection to the intellectual property of the creators/owners of these tools and modules. Legal actions can also be taken on the service if the service provides false or tempered outputs which deviate from the real results. These deviations can cause several ethical issues like gender bias, racism, and sexuality bias. All people are equal in the eyes of law and the Equality act 2010 [36] protects these rights of the people. The model is supposed to abide by these laws and produce results that are not biased in anyway.

### B. Automation

As discussed in the previous section, the main purpose of using MLOps is to automate the pipelines, to increase performance and speed of the service. But we need to make sure the level of automation required for the service. Previously we discussed in brief about the levels of automation that Microsoft and google provide. Here we are going to have a deeper look at the google maturity model as it is more often used in ML services. As discussed, this model is divided into three levels with increasing level of automation.

- **Level 0: Manual Process**

    Level 0 is a basic level of maturity and the teams associated with the process of model development and deployment do all the work manually. Figure A 1(top) shows the block diagram of a level 0 model using google maturity model. The data scientists create the models in isolation. Using notebooks, they write the code and test it multiple times running those notebooks. From training to evaluation, they do all the process manually. Even data analysis, preparation, pre-processing, and feature extraction tasks are manual. After the evaluation in multiple iterations, if the data scientist thinks that the performance of the model is good enough to be hosted on the internet, they hand over the notebooks to the Engineers team for the Operations task. This shows that there is least interaction involved between ML and Ops teams. The hand over can be in the form of uploading the model into a drive, repository, or a model registry. Then the engineers modify it to work in low-latency servers.

    There are a few drawbacks to this approach. The first being no or few version releases of the model. As everything is manual and creating a new version means doing everything manually from the start, every time. The only thing the team tends to do, to create a new version, is to retrain the model or change the whole model. This makes version updates very slow. These types of businesses release only a few versions per year. As there are less version updates, the use of CI and CD pipelines is ignored. In addition to this, the whole ML system is not deployed. The model is deployed using rest APIs only as a service to make predictions which usually are not logged, hence there is no way to check if the model is degrading. Such types of models often break after some time of deployment as they fail to adapt the changing conditions of the real world, and to prevent this, continuous monitoring and retraining are a must.

- **Level 1: ML pipeline automation**

    In level 1 automation, the ML pipeline is automated. Due to this the model keeps on training on new data continuously that helps the model to do predictions continuously. To support the structure shown in Figure A 1 (centre) and add automation to the model training process, some additional components are added. These include the data and model validation, feature and metadata store, and triggers. All these components are important and allow us to do fast experimentation, continuously train and deliver the model to make predictions, and store the ml model a container or module. In this way, we create a pipeline which becomes the product, in contrast to the Level 0, where only the model is the product. Owing to the importance of these components, we need to discuss each of them separately.

    **Data and model validation**

    Both data and model validation play a vital role in the automation process. When one or more triggers are activated, the ML pipeline starts to execute automatically, causing the model to retrain. The triggers can be one of the following:
    1) on demand
    2) scheduled
    3) new data available
    4) model performance degraded
    5) concept shift

    The pipeline requires new data to train the model on, and for that validation of the data is necessary so that the pipeline can decide whether to retrain the model on the coming data or stop the pipeline to create a new model. At data validation stage, the change in data could be of two types. Either the values of the new data are changed, or it is not compliant to the previous data scheme. In the first case, the pipeline would just retrain the model and release a new version. But in the latter case, the pipeline would stop its operation, because it means that the features of the new data are either partially or completely different from the previous training data and the model is not capable of predicting the correct results with this type of data. The data scientist than check the problem, create a solution, and then produce a new version of the model to start the service again.

    **Feature and metadata store**

    A feature store, as discussed previously, is a crucial part of automation and connects with both automated pipeline for training and the prediction service for inferencing, providing them with definitions, features values and storage. Adding feature stores to automation pipeline helps reuse the features again and again without requiring to create the features every time. The feature stores store up to date features, avoiding duplication of the features, to be used when required by the data scientists. In this way, feature stores are used by the data scientists for their experiments, by the service for predictions, and by the CT pipeline for re-training the model.

    In addition to the features, the metadata is also necessary for a smooth working and diagnosis of the pipeline. Storing metadata helps in debugging the problems as well as during reproduction of the pipeline. The pipeline stores the information of every execution, its date, time, steps involved, arguments passed, etc into a metadata store. It also stores the pointers to previous versions of the model in case a previous version is to be called and run in the pipeline. The evaluation results of each version are also stored in this storage and can be brought up if the pipeline performance is to be observed over time in different versions.

- **Level 2: CI/CD pipeline automation**

    The last level of automation in the Google maturity model is level 2. This model suitable if fast version updates are required and uses automation to the CI an CD pipelines for this purpose. The automation helps explore the data and generate new features more rapidly facilitating tuning of the model and hence give better performance at a higher deployment rate. The CI pipeline is used to build all the model components like the container images, executables and packages and are put through various tests where different logics and functions being

implemented in the model are tested. There are numerous tests which include model convergence, NaN values test, and the integrity of the components of the pipeline. Whereas the CD pipeline is used to continuously deliver the new trained model to the server so it can make predictions continuously. For the smooth running of the automation, it is important to verify that the packages and modules used are same everywhere. It is a good practice to load test and check the prediction service to see if giving a known input, does it predict the right output. The workflow of the level 2 maturity model is shown in Figure A 1 (bottom).

*C. Tools*

After all the discussion on MLOps workflow and its importance, there is a need to discuss the tools and platforms which are used to make these efficient pipelines that are automated to do all the tasks in the ML model life cycle from data exploration to hosting the model as a prediction service and introducing CI, CD and CT capabilities to these pipelines. There are a huge number of tools now available [37] in market for each task in the pipeline and these tools work in cooperation with each other to give the reasonable prediction results, speed, and performance. Some of these tools are available free of cost as opensource products, while some are public licenced, and some are for private use only. The open-source tools and platforms are used the most in the field of AI and MLOps and are estimated to be around two thirds of the whole community [38]. A big example of open-source tools use can be observed in the form of the developers available on GitHub [39] producing hundreds of millions of projects every year. GitHub itself, claimed that the community of developers on the platform has reached to around 100 million as of January 2023 in a recent blog post [40]. In this section, these tools will be discussed in detail, keeping in view the reason they are used for. Similarly, the versioning tools are used to create multiple versions of the data and stored in a well-organized way. Creating multiple versions help train models on different configurations of data to compare the performance of each version.

- **Data pre-processing tools.**

There are several tools that are available for performing pre-processing tasks. These tools are of two types. One of the types is called labelling tools and the other is called versioning tools. The labelling tools, as suggested by the name, are used to label the data. These could be for the annotation of image data, changing file names, specifying file types [41], etc. Different tools perform the tasks differently. In image processing, the labelling can be done using bounding boxes, but there is also the way of using semantic segmentation. The performance of the model is dependent on the tool that is used as it is somewhat based on the labelling accuracy [42] of the tool. Figure 5 shows some of the pre-processing tools available online. The list includes both open-source and private tools, and the type pre-processing they do.

- **Modelling tools**

Modelling tools are used for three purposes and hence have three types. The first type is featuring extraction tools which are used to do fast feature engineering on the dataset to find out the best features in the data set. The other type is the experiment tracking tool, which help keep track of various versions of the task created with different model and hyperparameters. And the third is used for hyperparameter tuning and are called hyperparameter optimization tools. Some industry famous tools are shown in Figure 6 below.

| Name | Status | Launch year | Tool type |
|---|---|---|---|
| Google Vizier | Public | 2017 | Hyperparameter Optimization |
| SigOpt | Public | 2014 | Hyperparameter Optimization |
| TensorBoard | Open-Source | 2017 | Experiment Tracking |
| Scikit-Optimize | Open-Source | 2017 | Hyperparameter Optimization |
| Hyperopt | Open-Source | 2013 | Hyperparameter Optimization |
| CML | Open-Source | 2017 | Experiment Tracking |
| MLFlow | Open-Source | 2018 | Experiment Tracking |
| Iguazio Data Science Platform | Private | 2014 | Feature Engineering |
| TsFresh | Private | 2016 | Feature Engineering |
| Featuretools | Private | 2017 | Feature Engineering |
| Comet | Private | 2017 | Experiment Tracking |
| Neptune.ai | Private | 2017 | Experiment Tracking |

Figure 6: Modelling Tools [38]

- **Operationalization tools**

The machine learning models when put to real world, they face degradation after some period of time as the data never remains the same and due to this data drift changes have to be made to the prediction service. For this reason, the model has to be monitored at all times. This task is usually done using tools that raise an alert in case there is a data drift or any anomalies in the pipeline. There are also tools that help integrate the ML models into production and are called deployment tools. In addition to these two, end-to-end tools are also available that are able to perform every step in the ML model life cycle. Figure 7 shows a few of these operationalization tools.

| Name | Status | Launch year | Tool type |
|---|---|---|---|
| Google Cloud Platform | Public | 2008 | end-to-end |
| Microsoft Azure | Public | 2010 | end-to-end |
| Databricks | Private | 2015 | end-to-end |
| TensorFlow Serving | Open-Source | 2016 | Model Deployment |
| Amazon SageMaker | Public | 2017 | end-to-end |
| Kubeflow | Open-Source | 2018 | Model Deployment |
| MLFlow | Open-Source | 2018 | end-to-end |
| DataRobot | Private | 2019 | end-to-end |
| Torch Serve | Open-Source | 2020 | Model Deployment |
| Arize | Private | 2020 | Model Monitoring |
| Deep checks | Private | 2021 | Model Monitoring |

Figure 7: Operationalization Tools [38]

As every task is different from the other, there is no set rule to define which tool is the best. The selection of tools is dependent on the application. The good approach is to use the least number of tools possible to give the desired performance as adding more tools can cause compatibility issues. But in case we need more flexibility, we can use more tools. S there is always a trade-off between compatibility and flexibility. To have a good practice with MLOps learning as many tools as possible is better so that we know which tools can be used to keep the balance between compatibility and flexibility.

IV. MLOPS FRAMEWORKS

Over the years, the need for MLOps frameworks has increased as more and more companies and businesses are moving towards industrial level machine learning services. These frameworks provide the users with special tools that help them through the whole ML service lifecycle. These frameworks are intended to handle machine learning's particular issues, such as data pre-processing, model construction, and model deployment. We will look into a few prominent MLOps frameworks in this section. These include TensorFlow Extended (TFX), Kubeflow and AWS SageMaker.

| Name | Status | Launch year | Tool type |
|---|---|---|---|
| iMerit | Private | 2012 | Data Pre-processing |
| Pachyderm | Private | 2014 | Data Versioning |
| Labelbox | Private | 2017 | Data Pre-processing |
| Prodigy | Private | 2017 | Data Pre-processing |
| Comet | private | 2017 | Data Versioning |
| Data Version Control | Open-Source | 2017 | Data Versioning |
| Qri | Open-Source | 2018 | Data Versioning |
| Weights and Biases | Private | 2018 | Data Versioning |
| Delta Lake | Open-Source | 2019 | Data Versioning |
| Doccano | Open-Source | 2019 | Data Pre-processing |
| Snorkel | Private | 2019 | Data Pre-processing |
| Supervisely | Private | 2019 | Data Pre-processing |
| Segments.ai | Private | 2020 | Data Pre-processing |
| Dolt | Open-Source | 2020 | Data Versioning |
| LakeFs | Open-Source | 2020 | Data Versioning |

Figure 5: Data pre-processing tools [38]

**TensorFlow Extended**

TFX is an open-source, end to end MLOps workflow which helps the users to produce scalable ML models [51]. This platform is produced by google and can be used to create custom MLOps pipelines as required by the user. As this is an end-to-end platform, it provides the user with all the tools that are necessary for any step in the MLOps workflow. There are tools available for data pre-processing, model development and deployment and provide help in creating and managing the pipelines which can be automated for Continuous Training, Continuous Integration and Continuous Delivery purposes. It supports a vast variety of libraries including all the famous and known ones like TensorFlow Hub, Model Optimization, and Tensor2Tensor, etc. [52].

One of the most important features of TFX is its scalability. It helps the user to produce models that are scalable and can be deployed on a variety of platforms. The user can use a deployment tool like Kubernetes to deploy these models with ease. It also helps track and manage the model after its deployment and provides the facility to validate and version the models as well. TFX contains an Evaluator [53] (Model Validator – Deprecated [54]) component, which allows users to evaluate the model's validity before it is deployed. It also has a Model Pusher component [55], which enables users to deploy the model to a production environment, such as a cloud-based platform or a mobile app.

**Kubeflow**

Kubeflow is an open-source platform for designing, deployment, and management of ML workflows that runs on Kubernetes [56]. Kubeflow is a set of tools that data scientists and developers may use to create, test, and deploy ML models. It offers a end to end platform for scalable ML models through every step in the MLOps Workflow. Just like TFX it also supports various known libraries. TensorFlow, PyTorch, and Apache MXNet are among the prominent machine learning frameworks supported by Kubeflow. It also includes a collection of pre-built components known as Kubeflow Pipelines, which may be used to create ML Workflows. Developers may use Kubeflow Pipelines to construct and manage ML processes as code, making it easy to version and recreate machine learning experiments [57]. A key benefit of using this platform is its integration with Kubernetes which makes it scalable and manages large data. The Kubeflow notebooks allow the user to make changes in the web-based process directly in the Kubernetes cluster making the task easier [58]. Another important component of Kubeflow is called katib which allows hyperparameter tuning and neural architecture search [59].

**SageMaker**

SageMaker is yet another platform that helps users to create, train and deploy ML services for custom problems providing them with specialized tools that suit their need [60]. This AWS based platform also has a no-code feature to help business analysts take advantage of this service [61]. A prominent feature of this service is AutoML [62] that automatically creates ML models based on the provided data with complete visibility and control. It selects the best model to suit the use case and easily deploys it with a single click. It also provides automatic hyperparameter tuning to choose the most efficient version of the model [63]. Moreover, it also provides with dozens of other features like feature stores [64], deployment [65], distribution [66], debugging [67], monitoring [68], and pipeline management [69].

V. FEATURE STORES

The popularity of ML models has greatly grown in the few last years due to their capacity to assess complicated data and anticipate outcomes. Nevertheless, in order to train and deploy these models, enormous volumes of data must be pre-processed to analyse and create the proper features. Feature stores evolved as a method for managing these complicated data pipelines by offering a single repository for storing, maintaining, and serving the features. Although, we have discussed in brief, about the feature stores previously, it is crucial that we discuss in detail its role and importance, along with its limitations, owing to its importance in the industrial level production of the ML projects.

Data, itself, is a very powerful tool and the features of this data are the foundation of this power. The right features for a task are, no doubt, important but are also hard to create. Feature engineering is therefore a critical part of an ML model lifecycle. Choosing the right features lead to creating a right model and hence good business outcomes. The features can be any sort of data, such as numerical, category, or text data, retrieved from a variety of sources, including databases, log files, and sensor data. The feature store assures consistency and reliability of the data, and it may be used by many teams to train and deliver models. Data versioning, data lineage, and data validation may also be handled by feature stores, ensuring that the data is correct and up to date.

Data collection, validation and pre-processing is time consuming and takes too much effort. Creating new features is complicated and uses hit and trail method until you are completely satisfied with your one-of-a-kind new feature. Then you have to compute and save it as a production pipeline, which varies based on whether the feature is online or offline. The introduction of feature stores completely resolved this issue letting the data scientists work on actual model building rather than spending enormous amount of time on feature creation and data pre-processing. A main issue with the traditional method is that the data is not hosted on a single platform and the data scientist has to look for it at various places to gather relevant features. Feature stores help them catalogue the features at a single platform where they can quickly discover new feature and use them as they are for any ML model.

Feature stores are very important in scaling ML applications. As ML models get more complicated, enormous volumes of data are required to train and evaluate them. Feature stores allow teams to easily handle huge datasets while offering a single point of information for all features used in models. This strategy avoids duplication of labour, assures data consistency, and boosts the overall productivity. Moreover, the feature stores help in deploying the model into production as they manage all the models by providing them a single repository which they can utilise and access at all times.

So, it can easily be said that the features stores provide multiple benefits for ML pipelines. They increase the productivity of the team by letting them spend more time on model development than the pre-processing task. It increases scalability and help in efficient production of the pipelines. By providing a single platform to access the features, it increases collaboration between multiple teams. And most importantly, it increases the performance of the model by providing consistent and reliable features that are reusable as well.

## VI. Design an Enterprise ML System

In this section, we will be using the concepts of MLOps and implement them to deploy a machine learning prediction service. We will be working on an object detection service using TensorFlow 2 and some other tools. For this example, we are not going to create a custom object detection model, but instead use a pre-trained model from TensorFlow model zoo. This example task is just to give an idea of how MLOPs can be used to host real time prediction models and the concepts can be implemented on any type of model.

**Requirements**

Before starting on the model workflow and implementation, we need to discuss all the requirements for this task. The most important things that we need for this task are as follows:

- Machine Learning model
- Web app to provide the ML service
- Tool/Platform to package the model
- Tool/Platform to Deploy the service
- Model monitoring
- Automatic CI - CD pipeline

The model that we are going to use will be an SSD ResNet 101 model and will act as the first version of the model. We are not going to train the model from scratch but use a frozen version from the TF2 model zoo [43]. To create a web app, we will use flask and create a python application and will package the model with the web app using a Docker [44] container image. To deploy this package, we will use a hosting platform like Heroku [45] with GitHub actions [46] or Kubernetes [47] where we will host the model and monitor it as well. In case of a version upgrade, we can push a new TF2 object detection model, say Faster RCNN ResNet 101, to the Kubernetes cluster and deploy it as a new version. Let's have a brief discussion on these tools before we get to the workflow of the model.

The first tool that we require is Docker. Docker is an open-source platform that allows the user to create virtual environments called containers that are completely isolated and independent of other file systems. These containers have their independent dependencies, libraries and tools and share just only the host device's OS kernel. The container is like a separate isolated room in a house where the user can perform a specific task without any sort of disturbance from outside that environment. A great benefit of containerizing you work is that you can take the container to any system, and it will work the same as it has no link to the outside dependencies and has its own dependencies inside its environment. This makes the containers portable, efficient, consistent, and secure, along with high-speed deployment. Because of these properties, using containers is a more popular way of application development and deployment. Docker containers, although are the most common and popular in the market there are other options like LXD and LXC [48] available that provide containerized solutions.

For automated CI – CD, we need a tool that could automatically build, test, and deploy the service on the internet. GitHub actions can be used for this purpose. It can be used to build and test all the pull request to the GitHub repository and deploy these to the servers. GitHub actions is popular as it provides the user with Virtual Machines (VMs) for Linux, MAC OS, and Windows, along with the option to self-host their data through data centres and cloud services [49]. When a repository event happens, such as the opening of a pull request or the creation of an issue, the GitHub Actions procedure is initiated. The workflow is made up of one or more jobs that might run sequentially or simultaneously. Each job will run in its own VM runner or container and in a few steps it will either execute a script or an action, which is a reusable feature that may help you streamline your workflow.

To host the model on the internet, we need a cloud platform that support our model. Heroku is a very good option and works great with GitHub actions. User has to provide some secret keys to connect the two platforms and then can push the service on the server. Another option is Kubernetes, which is very popular now and is a service by Google. It is an open-source system and can manage containerized apps. Its purpose is to manage the whole lifecycle of the containerized apps. Kubernetes is used for Enterprise scale ML projects and helps maintain multiple containers at a time. Sometimes, Docker and Kubernetes are confused with each other. To make things simple, we can say Dockers is used to make containers while Kubernetes is used to manage these containers allowing them to work together and reduce the workload. Docker is used in the lifecycle of any programme to package the application at the moment of deployment, whereas Kubernetes is used for the rest of the life to maintain the application. Kubernetes scales, runs, and monitor the service. Figure 8: Lifecycle of an App *[50]* explains the same thing.

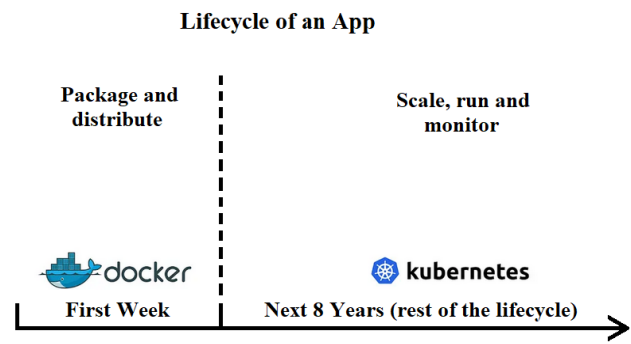

Figure 8: Lifecycle of an App [50]

**System Design**

The ML prediction service system design starts by getting an understanding of the business model and collecting and validating the data required for it. In this case, we want to design a web app that would be capable of detecting objects from an image uploaded to it. To train this type of model, we need a huge image data set, validate these images, and then label each object in all the test images set. After this step the model is trained on these test images to be able to make predictions. To keep things simple, we are not going to collect the images or label them. Instead, we will use a pre-trained model from TF2 model zoo. The models present on the model zoo are trained using huge datasets of labelled images and on super-fast and expensive equipment. The models available, the frozen versions of these trained models. We can use these pre-trained models to create our own object detection service.

**ML model**

The architecture that we will use for this task is an SSD ResNet 101 architecture. The model has been trained using the COCO dataset [70] and gives very good performance. The mean average precision for this model on COCO data set is reported to be 39.5 which is a fairly good value as compared to the other models available on the model zoo. The inference time is 104 ms which is a little higher. There are other models that are better performing than this one, but we have chosen this model just to depict how a model is used in enterprise ML services. This model will serve as out first version, and we will deploy a better version later as a version update.

Similarly, if we wanted to train our own custom object detection model, we would have to take a few more steps required for the training purpose. These steps include all, starting from data collection and validation till the model training and evaluation, so that we have a trained model that we can host. Next few passages discuss these steps briefly to give an idea on training our own models.

- **Data collection and validation**

The first step in model training process is to collect the data. The data comes from various sources and is in the form of images. These images can range from a few hundreds to tens of thousands depending on the availability of the data. Once the images are collected, it is important to make sure that all the images are valid and usable. Checking the data is very important as we do not want our model to train on faulty images. The data scientists make sure that the data is usable and is in the form it is supposed to be in. For example, if we are training a model for the detection of birds, we need to make sure, all the images are sorted correctly and include the right species of birds we intend to classify. Image resolution and quality also matters when training. Poor quality images are hard to extract features from. These types of images should be avoided.

- **Pre-Processing and Data labelling**

The next step in this process is to pre-process the data. The data is split into partitions for testing and training as we cannot use the same images for both. The properties of the data set are observed and important manipulations like data augmentation is performed during pre-processing. Image augmentation is very helpful in cases where we have very little data set. Applying various manipulations on the test images help the model train on all the features of the data set as it gets a chance to observe the images from various perspectives. For example, we can perform rotation, zoom, contrast etc. ImgAug [71] is a good module for image augmentation, as it takes labelled images as input and provide augmented images along with augmented labels saving the user from the hassle of re-labelling the images.

Labelling/tagging the data is crucial as we need our model to understand which object has what name. There are a lot of modules available open source which can be leveraged to label our images. A good example of such a module is 'LabelImg' [72]. It is an easy-to-use module and can store the labelled data in any form that we require. Figure 9 shows an image being labelled.

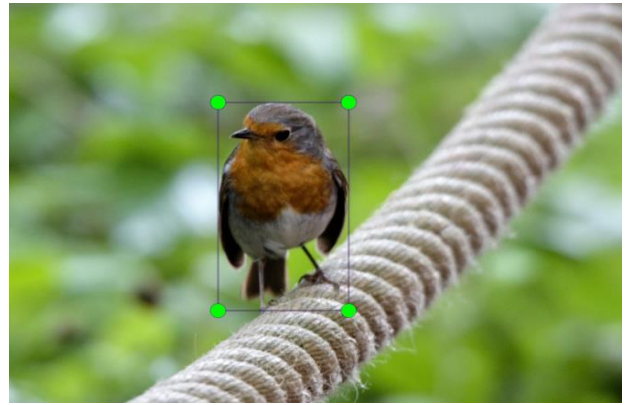

Figure 9: Image Labeling

- **Framework selection**

After we have the dataset pre-processed and labelled, we can move on towards the training process. For the training, a good option is to use transfer learning technique. In this method, we can leverage an already pre trained model, like yolo or any other model from TF Model Zoo and train it for specifically for our custom objects. Training a model from scratch requires extremely huge data sets and computational resources that are not available most of the times. The option to use the capabilities of a pre-trained model to create custom models is a blessing in such cases. We can use the model that meets our requirements from TF model zoo and train it on our custom data. Figure 10 shows a list of some famous models and their stats as reported by TF model zoo.

| Method | Comparison of Inference time of YOLOv3, SSD512 and Faster R-CNN | | |
|---|---|---|---|
| | mAP | Inference time (ms) | Frames per second (FPS) |
| YOLOv3-320 | 51.5 | 22 | 45 |
| YOLOv3-416 | 55.3 | 29 | 34 |
| YOLOv3-608 | 57.9 | 51 | 19 |
| SSD512 | 76.8 | 45 | 22 |
| Faster R-CNN | 34.7 | 37.7 | 26 |

Figure 10: Inference time and mAP comparison

- **Evaluation and Hyper-parameter tuning**

For a trained model to perform good in the lab and collapse in real world environment is very common. To make sure the model will work as intended in the place where it is supposed to be deployed, we need to make sure that it meets the performance criteria specified. TensorFlow provides the option of model monitoring during and after training using the Tensor board. There are various metrics that can be checked to evaluate the model. These metrics are based on all the prediction made by the model. The predictions could be of any four types: True Positive, False

Positive, True Negative, and False Negative. Tensor board directly displays mean Average Precision (mAP) and Average Recall (AR), at different image sizes and different number of objects per image, in the form of plots. These metrics decide the performance of the model based on what application it is trained for. To increase the performance, techniques like hyperparameter tuning are used where the hyperparameters are changed to train the model until the required performance is achieved. A few important hyperparameters are the learning rate, batch size and number of epochs. Playing around with these parameters can help increase the performance of the model. Figure 11 shows a comparison of the performance of a few models as reported by their research papers.

| Method | Performance comparison of YOLOv3, SSD512 and Faster R-CNN trained with MS COCO trainval35k dataset | | | | | |
|---|---|---|---|---|---|---|
| | $AP$ | $AP_{IoU}$ | | $AP_{Size}$ | | |
| | | $AP_{0.50}$ | $AP_{0.75}$ | $AP_L$ | $AP_M$ | $AP_S$ |
| YOLOv3 | 33.0 | 57.9 | 34.4 | 41.9 | 35.4 | 18.3 |
| SSD512 | 26.8 | 27.8 | 46.5 | 41.9 | 28.9 | 9.0 |
| Faster R-CNN | 34.7 | 55.5 | 36.7 | 52.0 | 38.1 | 13.5 |

Figure 11: Performance comparison

**Web application development**

After the model is in a working state, and giving predictions, we need to create a web app to provide an interface for the users where they can upload the images and get the predictions of the objects in the images. We will be creating the app using Flask and HTML and add styles and designs to it using CSS. The app will consist of multiple webpages. The home page of the web app will give an intro the application. A separate page will have the option to input the images. The output will be shown in a separate window as the image with the detections made on them. TF Serving is used to serve the model to get back predictions using either gRPC or REST API. One of the other webpages will be the about us page, which gives an overview of the company, its goal, and achievements. There will also be a feedback page, where the user can use the provided form to give feedback to the service by leaving their name and email with a message. A contact-us page will also be provided in case if the user wants to contact and report anything. More on the project can be found in the GitHub repository for this project [76].

**Package the model**

After the creation of the web app and the ML model to be hosted, we will require to create a docker container to package this model. For this purpose, we create a docker file. A docker file is foundation of creating a build using docker and is used to create and manage an image of the build. The file consists of commands that are required to instruct docker on how to build the image, install dependencies and run the package. An image is a light-weight package that contains all the required libraires, codes and dependencies so that the build is reproducible and consistent in all the environments. Figure 12 shows a docker file that we can use to create the image of our model. This is a multi-stage docker file that has two stages. The first stage (build-stage) will create an intermediate image, which will copy the flask app and install the dependencies. The second stage (runtime-stage) will create a smaller image by copying the libraries and flask app from the first image to this image. Multi-stage docker files are very beneficial if we want to create small sized images as it reduces the image size to a great extent. A noticeable thing in the docker file is the use of python instead of a full operational operating system as out base images. There are multiple benefits of taking this approach. A few of these include reduced image size, simplified deployment, efficient resource utilization, and faster build times. Moreover, we have used 'RUN' only once while installing the packages. If 'RUN' is used more than one time, multiple layers are created, which require more memory as compared to using it only once. A number of tags are used when installing the packages which are explained below:

- '--no-install-recommends' in apt-get command stops recommended installations and installs only specified libraries saving memory.
- '--no-cache-dir' with pip command stops the cache directories to be downloaded saving memory.
- '--disable-pip-version-check' with pip command stops the console to check the version of pip, saving time.

```
# Build stage: (Base image is python 3.7 that is on a light weight linux platform)
FROM python:3.7 AS build-stage

# Set the work directory
WORKDIR /app

# Add project
ADD FlaskObjectDetection /app/FlaskObjectDetection

# Download and install dependencies (Using RUN only once reduces layers and saves memory)
RUN apt-get update && \
    apt-get install --no-install-recommends -y ffmpeg libsm6 libxext6 && \
    pip install --no-cache-dir --disable-pip-version-check Flask tensorflow \
    tensorflow-serving-api mysqlclient Flask_mysqldb numpy Pillow matplotlib grpcio pandas protobuf==3.20.*

# Runtime stage: (Base image is python 3.7 that is on a light weight linux platform)
FROM python:3.7 AS runtime-stage

# Set the work directory
WORKDIR /app
# Copy the site packages from build image to run time image
COPY --from=build-stage /usr/local/lib/python3.7/site-packages /usr/local/lib/python3.7/site-packages
# copy the flask app from build image to runtime image
COPY --from=build-stage /app/FlaskObjectDetection /app/FlaskObjectDetection
# Set environment variables
# PYTHONUNBUFFERED=1 lets the output display immediately instead of storing it in a buffer with takes more time.
ENV PYTHONUNBUFFERED=1
# Tell the image what to do when it starts as container
WORKDIR /app/FlaskObjectDetection
# in command promt run the flask app.py
CMD ["python", "./app.py"]
```

Figure 12: Docker file

**Deployment and monitoring**

The container that we created now needs to be deployed to the server for making predictions. We will take two examples for deployment and the workflows are explained in the Figure 13 a and Figure 13 b. The workflow in Figure 13 a will be using Kubernetes as a means of deployment and monitoring the model. We will create a new project in Kubernetes and bring our container to the platform. We will have to authenticate the docker container to the container registry first, but it is a one-time process only. After we have our container on Kubernetes, we can create a cluster and deploy the service to the server and expose it to the internet. A cluster is made up of a collection of Compute Engine VM instances that run Kubernetes.

At this point we would have successfully deployed the model to the internet but, this is not the end of the life cycle of the ML service. The ML models are not like basic Dev services, and they start degrading as soon as they come in contact with the real-world inputs. So, over the time, the model will start to lose its performance, and we need to keep an eye on this degradation. We would want to monitor the model so that we know if there is a change in trends of the data of if there is a complete data shift, that could cause model to stop working properly. In that case, we need to change the model to some extent, to cope with the changes or sometimes, if the changes are drastic, we change the architecture completely. There could be numerous triggers that would suggest making changes in the model as we discussed in a previous section. For example, we need to update the model for some reason, we will require a new model that has better performance than the previous version, for our case, it could be Faster RCNN ResNet 101 with 37.1% mean average precision on COCO dataset and an inference time of 72 ms making it similarly precise and a lot faster than the SSD model. We can upload this model to our cluster using Kubernetes providing us with continuous integration and continuous deployment in the pipeline.

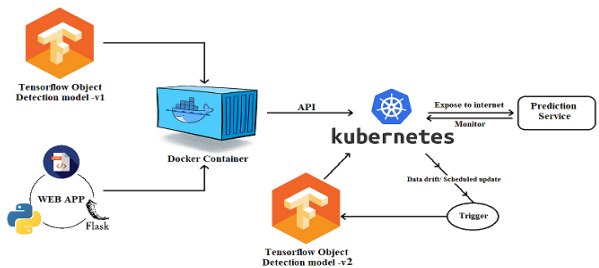

Figure 13 a: Workflow using Docker and Kubernetes

Kubernetes is a paid service and charges the user for deploying model on the server. This platform is usually used in advanced Enterprise level ML service models. A cheaper way to do for new starters is using Heroku as method of deployment where it is usually free to use the service to learn and understand as they give weekly free hours to the users. But when you need to deploy your business as an online service, you should go for investing in a good tool as it will pay off better than using a cheap or free tool that has low performance.

Heroku uses GitHub actions to push the model to the server. The user uploads the model, containerised, or non-containerised in a GitHub repository, and link the repository using secret action keys with Heroku, giving it access to the repository. After this Heroku can serve the model online. The changes to the model can be made using the GitHub repository. The new updated version of the model is uploaded to the GitHub repository and the Heroku server uses GitHub actions as a CI- CD pipeline to deploy the model to the server. The workflow of the ML service using GitHub and Heroku is shown in Figure 13 b**Error! Reference source not found.**.

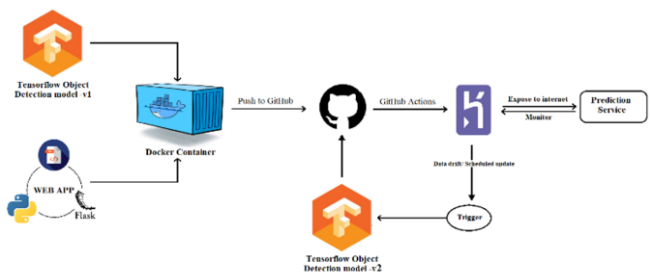

Figure 13 b: Workflow using GitHub Actions and Heroku

**Considerations for ML system design**

**Containerisation/ Non-containerisation**

A major consideration for this task is to use containerised pipelines. We can also use non-containerised solutions too, but it adds to the complexity of the task and can cause numerous errors to arise. A non-containerized MLOps pipeline often entails manually installing and configuring the important software necessary for training, testing, and deploying machine learning models. This requires the user to install all correct versions of the dependencies, libraries and frameworks, maintaining software components, and customising runtime environments. Such pipelines can be difficult to set up and manage, and they can be susceptible to compatibility problems and versioning disputes. In contrast to this, a containerised pipeline creates an isolated environment, where each and every dependency is already available, and can cause no issues related to version dispute hence running the model/ source code smoothly as it is intended to be. Creating a container helps the other teams from getting rid of the errors and delays caused because of version miss match and the programs run exactly the same as they run on the computer system they were created on. Thus, using Containerised solutions create smooth collaboration and sharing between various teams involved in production task. Containerisation also allows for more efficient resource use because containers may be dynamically spun up and down based on workload needs. This implies that developers can simply grow their ML pipelines without having to worry about manually managing and procuring hardware resources.

**Roll-back and versioning**

It is crucial to consider the importance of versioning in ML model production and deployment. Keeping the log of models released helps the business in the long run. Versioning can be a small update in the code or a completely new model to overcome the data drift. In both cases it is important to track the versions. There can be various ways that a model can be versioned while in production. Separate containers can be created for different versions of the model and the new container can be hosted as a new version. GitHub Actions is a commonly used Version Control System that help store and manage various versions of the ML model. The containers can also be hosted on Kubernetes cluster helping in the management of the versions using YAML files. Kubernetes also has a rolling update feature, and utilising this capability helps deliver new versions of your ML models while minimising the downtime. It updates a subset of replicas at a time, allowing you to progressively move to the new version while ensuring that it is operationally sound. In case, the previous version was a better option, Kubernetes also has a roll-back feature that helps create a new version of the ML model by reusing the previous version configuration hence, restoring it.

**Infrastructure test**

An infrastructure test plan is a necessary component of any ML service that strives for accuracy, scalability, and high availability. It contributes to ensuring that the system can manage increasing traffic and load while being robust to breakdowns. There are multiple tests that the system should pass before it can be put to service. We will be performing these tests on our prediction service to make sure it performs as expected in the real-world environment where various types of traffic will be introduced to it at all times and the performance does not degrade before the expected time.

Following are a few of these tests that are a must for this task to be a success.

- Load test
- Performance test
- Fault and latency test (Chaos Engineering)
- Rolling Back and Replacing Models
- Disaster Recovery

**Load test**

Load testing [73] is an important part of the test phase which ensures that the system (model and the device it is running on) can withstand increased traffic and load. There are a set of service level agreements related to every project and these standards are to be met for the model to be considered as well performing model. Load testing gives the idea of the capability of the model to handle load and at what amount of traffic it will fail so that this information can be used to create better versions of the model in future. A platform like Apache JMeter [74] can be really helpful to assess the capacity of the model towards the incoming traffic. It is an open-source based on java and is widely used for the purpose of load testing. Using this platform, we can perform the following load tests easily to assess our service.

- A- *Baseline test:* The purpose of the baseline test is to determine the system's baseline performance under regular traffic situations. Over a set amount of time, the system will be evaluated with a low traffic load. During this time period, the performance metrics like response time and throughput are recorded. With these parameters, we know where the system stands in case of baseline traffic.
- B- *High traffic test:* This test is the opposite to the baseline test and makes sure that the system can handle a high traffic load. To perform this test, high traffic is introduced to the system for a time period and the same metrics, response time and throughput, are observed. If the system withstands the required standard for high traffic, this test is considered as a success.
- C- *Burst traffic test:* As the name suggests, this test will introduce bursts of high traffic at various instances, creating spikes in the traffic and measure the related performance metrics to ensure that the model will work properly if there is a sudden increase in the traffic for short period of time.

**Performance testing**

One of the most crucial parts of an ML service launch is the testing its performance. With this, we want to make sure that the application is stable under varying loads, works at the specified speed as business applications usually require models to give predictions in seconds, and scalable o the maximum required number of users at a time. We do performance test so that we have an idea of what changes are to be made to the model if it does not meet the required specifications. Apache JMeter, in addition to the load tests, can also perform performance test and proves to be a great tool in this regard. There are multiple tests involved in the performance test. A few of these tests that we will perform are discussed below.

- A- *Response time test*: Response time test is performed to see the time it takes the system to give predictions under various load parameters. It helps determine if the system will be able to cope up with the load while maintaining a suitable response time at the same time.
- B- *Concurrency test*: If the system is faced with multiple requests simultaneously, it might crash. To negate this possibility, we need to test it beforehand so that necessary measures could be taken in case such a situation arise. For this purpose, a concurrency test is performed which records performance metrics in a situation where multiple requests are sent to test the system.
- C- *Error rate test:* To make sure that our system is capable of handling errors while serving, we will perform the error rate test in which we introduce errors deliberately to the inputs and observe the performance of the system.

**Fault and latency test (Chaos Engineering)**

Performing this test, we will record the strength/ weakness of our system when critical faults are introduced. This test will show how resilient our system is and is also referred to as chaos engineering [75]. In this test we will introduce deliberate faults like network failure, disk failure, and server failure to our system and test its capability to handle these issues. We will also see if the system can handle latency issues as well.

**Rolling back and replacing models**

As discussed previously, we might need to replace or roll back to previous version of the model in case of data drift or outdated model. As we have automated pipelines for Continuous Delivery of the model, we need to test these before we deploy the model. It is done to make sure that the system works seamlessly when in a real-world scenario, we have to change our model while the current is still serving. For this purpose, we test the system by rolling back and replacing the model to simulate the situation.

**Disaster recovery test**

This is another important test that needs to be performed before deployment. We need to test the system for automatic backup and recovery in case the system breaks down because of a failure. This makes sure that the system will restore itself to a previous state in a catastrophic failure.

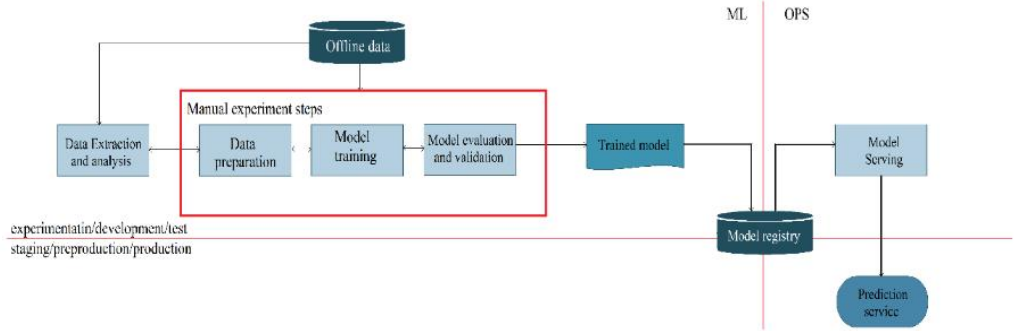
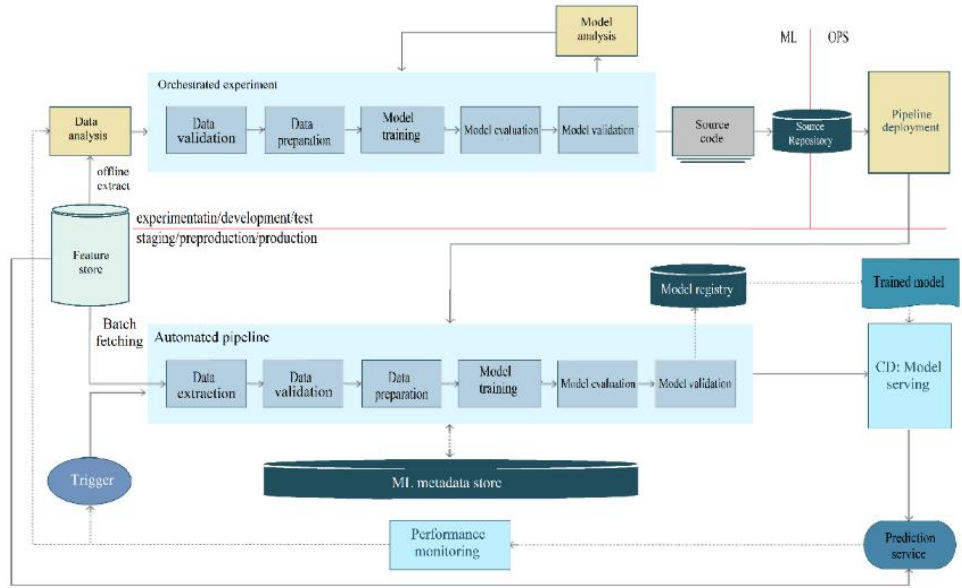
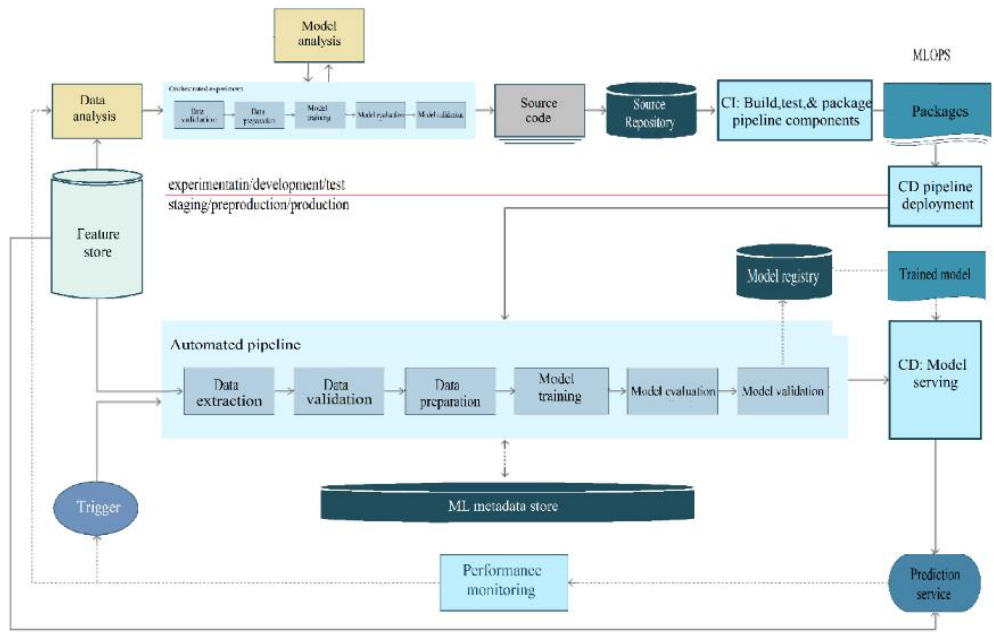

Figure A 1: Google maturity model - level 0, 1, and 2 [32]